# Most-Likely DCF Estimates of Magnetic Field Strength

Philip C. Myers[1], Ian W. Stephens[2], Simon Coudé[2]

1. Center for Astrophysics | Harvard and Smithsonian, 60 Garden Street, Cambridge MA 02138 USA; pmyers@cfa.harvard.edu

2. Department of Earth, Environment, and Physics, Worcester State University, Worcester, MA 01602, USA

## Abstract

The Davis-Chandrasekhar-Fermi (DCF) method is widely used to evaluate magnetic fields in star-forming regions. Yet it remains unclear how well DCF equations estimate the mean plane-of-the-sky field strength in a map region. To address this question, five DCF equations are applied to an idealized cloud map. Its polarization angles have a normal distribution with dispersion $\sigma_\theta$, and its density and velocity dispersion have negligible variation. Each DCF equation specifies a global field strength $B_{\rm DCF}$ and a distribution of local DCF estimates. The "most-likely" DCF field strength $B_{\rm ml}$ is the distribution mode (Chen et al. 2022), for which a correction factor $\beta_{\rm ml} \equiv B_{\rm ml}/B_{\rm DCF}$ is calculated analytically. For each equation $\beta_{\rm ml} < 1$, indicating that $B_{\rm DCF}$ is a biased estimator of $B_{\rm ml}$. The values of $\beta_{\rm ml}$ are $\beta_{\rm ml} \approx 0.7$ when $B_{\rm DCF} \propto \sigma_\theta^{-1}$ due to turbulent excitation of Afvénic MHD waves, and $\beta_{\rm ml} \approx 0.9$ when $B_{\rm DCF} \propto \sigma_\theta^{-1/2}$ due to non-Alfvénic MHD waves. These statistical correction factors $\beta_{\rm ml}$ have partial agreement with correction factors $\beta_{\rm sim}$ obtained from MHD simulations. The relative importance of the statistical correction is estimated by assuming that each simulation correction has both a statistical and a physical component. Then the standard, structure function, and original DCF equations appear most accurate because they require the least physical correction. Their relative physical correction factors are 0.1, 0.3, and 0.4 on a scale from 0 to 1. In contrast the large-angle and parallel-$\delta B$ equations have physical correction factors 0.6 and 0.7. These results may be useful in selecting DCF equations, within model limitations.

## 1. Introduction

Estimates of the magnetic field strength in star-forming clouds are used to assess their star-forming potential, by comparing their magnetic, gravitational, and turbulent energies. The most widely-used method is the statistical interpretation of polarization maps, assuming turbulent excitation of small-amplitude MHD waves (Davis 1951; Chandrasekhar & Fermi 1953, hereafter DCF; Ostriker et al. 2001, hereafter OSG01; Hildebrand et al. 2009; Houde et al. 2009, 2016; Pattle & Fissel 2019, Liu et al. 2021, Skalidis & Tassis 2021, Li et al. 2022, Chen et al. 2022, hereafter C22; Liu et al. 2022, Lazarian et al. 2022). Other methods of estimating field strength from observations include analyses of the Zeeman effect (Crutcher 2012), of velocity gradients (González-Casanova & Lazarian 2017) and of intensity gradients (Koch et al. 2012). These methods lie outside the scope of this paper, but they are discussed in recent reviews (Crutcher & Kemble 2019; Pattle & Fissel 2019).

Many authors have applied the DCF method to observations of star-forming regions, and several modifications to the basic DCF equation have been suggested (Liu et al. 2022, Li et al. 2022, C22.) It is unclear whether one version of the DCF equation differs significantly in its predictions from the others, and whether one version agrees significantly more closely with simulations than the others. Several studies have used simulations of MHD turbulent clouds to estimate "correction factors" to improve the accuracy of particular DCF equations. They find that DCF equations based on excitation of Alfvén waves by turbulent motions perpendicular to the mean field direction tend to overestimate the field strength, i.e. their correction factors have value $< 1$. In contrast a DCF equation based on non-Alfvén MHD waves driven parallel to the mean field slightly underestimates the field strength, i.e. it has correction factor $\gtrsim 1$ (Skalidis & Tassis 2021).

DCF overestimation of the field strength may arise for several physical and geometrical reasons. The wave magnetic energy may be less than the wave kinetic energy, contrary to the equipartition assumption of the DCF method (Heitsch et al. 2001, OSG01, Falceta-Gonçalves et al. 2008, Liu et al. 2021, Li et al. 2022, C22). The field strength may also be overestimated because the turbulence is anisotropic rather than isotropic (Skalidis et al. 2021, Lazarian et al. 2022). A similar overestimate may occur due to signal integration along or across the beam (Zweibel 1990, Myers & Goodman 1991, Houde et al. 2009), or because 3D angles may be reduced when they project onto the 2D plane of the sky (C22). Overestimation may also result from other angle approximations in deriving the DCF equation (Li et al. 2022), or because a cloud may have ordered field structure, due e.g. to cloud self-gravity and flux freezing (Mestel 1966). It is unclear which of these effects is most important, and whether their relative importance is similar from one DCF equation to the next, or from one simulation to the next.

Independent of these physical and geometrical effects, the quality of each DCF estimate may be improved by a probabilistic evaluation of the DCF equation. The DCF equation can be written in its usual "global" form or in a "local" form where each variable is a local analog of the corresponding global variable. Here the term "$B_{\mathrm{DCF}}$" is used as a short name for "$B_{\mathrm{DCF,global}}$." Thus the "original" DCF equation is $B_{\mathrm{DCF}} = (4\pi\rho)^{1/2}\,\sigma_{\mathrm{v}}/\sigma_{\theta}$, where $\rho$ and $\sigma_{\mathrm{v}}$ are mean densities and mean velocity dispersions over the map. Its local version is $B_{\mathrm{DCF,local}} = (4\pi\rho_{\mathrm{local}})^{1/2}\,\sigma_{\mathrm{v,local}}/|\theta|$, where $\rho_{\mathrm{local}}$ and $\sigma_{\mathrm{v,local}}$ are local estimates of density and velocity dispersion, each based on observation in a single telescope beam. Here $\theta \equiv \psi - \bar{\psi}$ is the local deviation of polarization angle $\psi$ from its mean $\bar{\psi}$ over the map. Then the most-likely estimate of $B_{\mathrm{DCF}}$ is the mode of the distribution of $B_{\mathrm{DCF,local}}$. This technique was applied to an extensive suite of MHD simulations (C22).

It should be emphasized that the most-likely field strength estimated from a particular DCF equation is the statistically most probable evaluation of that equation. This most-likely estimate is not necessarily more accurate than the most-likely estimate based on a different DCF equation. For that purpose each most-likely estimate is compared to the results of numerical simulations, as in Section 3.

In this paper, such "most-likely" estimates of DCF field strength are made analytically rather than numerically, for a simplified cloud model. The plane-of-the-sky angle deviations $\theta$ follow a normal distribution with dispersion $\sigma_{\theta}$. This assumption is supported by simulation results which show that angle distributions are well fit by Gaussian functions for $\sigma_{\theta} \lesssim 30°$ (OSG01). The value of $\rho_{\mathrm{local}}^{1/2}\sigma_{\mathrm{v,local}}$ in the numerator of the local DCF equation is assumed here to be negligibly different from the value of $\rho^{1/2}\sigma_{\mathrm{v}}$ in the numerator of the global DCF equation. This assumption is made to simplify the analysis and to minimize the number of parameters. It may correspond to a uniform distribution of $\rho_{\mathrm{local}}^{1/2}\sigma_{\mathrm{v,local}}$, or to a distribution whose spatial fluctuations are relatively small and uncorrelated with the variations in polarization angle. These conditions may be met for cloud regions close to virial equilibrium with relatively little stellar feedback, as discussed in Section 4.2.

With these choices, the mean polarization angle $\bar{\psi}$ is set to zero and the polarization angle dispersion $\sigma_\theta$ is the only free parameter. Although this approach is idealized, it allows direct comparison among DCF equations having the same value of $\sigma_\theta$.

The ratio of local and global field strength estimates for each DCF equation is here denoted $\beta \equiv B_{\mathrm{DCF,local}}/B_{\mathrm{DCF}}$. The distribution $p(\beta)$ is derived from the probability distribution of the deviation of polarization angles $p(\theta)$ via the change of variables relation (Papoulis 1965). The distribution $p(\beta)$ is then a normalized analytic version of the distributions $p(B_{\mathrm{DCF,local}})$ in C22 Figure 11. If $p(\beta)$ has a well-defined mode, the modal value of $\beta$ is denoted as $\beta_{\mathrm{ml}}$, the most-likely value of $\beta$ in the distribution. Then $\beta_{\mathrm{ml}} \equiv B_{\mathrm{ml}}/B_{\mathrm{DCF}}$ can be considered a correction factor of statistical origin used to obtain $B_{\mathrm{ml}}$ from $B_{\mathrm{DCF}}$.

This "most-likely correction factor" $\beta_{\mathrm{ml}}$ differs from the "simulation correction factor" $\beta_{\mathrm{sim}}$ derived from a numerical simulation of a MHD turbulent cloud. The simulation correction factor is defined as in OSG01 as $\beta_{\mathrm{sim}} \equiv B_{\mathrm{sim}}/B_{\mathrm{DCF}}$, where $B_{\mathrm{sim}}$ is the POS field strength in a region of a simulated cloud and where $B_{\mathrm{DCF}}$ is the estimate from the DCF equation using parameters from across the same region. The departure of $B_{\mathrm{DCF}}$ from $B_{\mathrm{sim}}$ is usually ascribed to physical properties and approximations, as discussed above. In this paper the departure of $B_{\mathrm{DCF}}$ from $B_{\mathrm{sim}}$ is ascribed to both statistical and physical properties. Five DCF equations are compared to identify which equations have the closest agreement between $\beta_{\mathrm{ml}}$ and $\beta_{\mathrm{sim}}$.

The plane-of-the-sky polarization angles are assumed to follow a Gaussian distribution with zero mean and standard deviation $\sigma_\theta$. This 2D standard normal distribution follows from a 3D standard normal distribution, only for sufficiently small values of angle dispersion, of mean field inclination, and for a sufficiently smooth variation of angles along the line of sight (C22 Figure 17).

In Section 2, this paper describes five well-known DCF equations**.** It presents expressions for their global versions in Section 2.1-2.5. It presents modifications to obtain their local versions in Table 1. Section 3 gives analytic expressions for their corresponding probability distributions $p(\beta)$ and for their statistical correction factors $\beta_{\mathrm{ml}}$. It compares the values of $\beta_{\mathrm{ml}}$ with each other, and with correction factors $\beta_{\mathrm{sim}}$ derived from simulations of MHD turbulent clouds. Three DCF equations have close agreement between $\beta_{\mathrm{ml}}$ and $\beta_{\mathrm{sim}}$, while two disagree more, and require significant non-statistical correction. Section 4 discusses these results and their implications, and Section 5 presents a summary and conclusions.

## 2. DCF Estimators of the Plane-of-the-Sky Field Strength

This section summarizes the five DCF equations considered in this paper, following their descriptions and names in their original papers, in Li et al. (2022), and in C22. A goal of this work is to compute a "most-likely" correction factor $\beta_{\rm ml}$ appropriate to each equation. Therefore each DCF equation is written here in its original form without any correction factor determined from previous studies. Its most-likely version is $B_{\rm ml} = \beta_{\rm ml} B_{\rm DCF}$ where $\beta_{\rm ml}$ is calculated from the appropriate distribution in Section 3 and is listed in Table 1. For example the field strength according to the original DCF equation is given in Section 2.1 below as $B_{\rm DCF,o}$ and its most-likely version is $B_{\rm ml,o} = \beta_{\rm ml,o} B_{\rm DCF,o} = 2^{-1/2} B_{\rm DCF,o}$.

The equations analyzed are selected based on their frequent use and on their simple dependence on $\rho$, $\sigma_v$, $\sigma_\theta$, with no additional parameters. Thus the structure function equation of Hildebrand et al. (2009) is included but those of Houde et al. (2009, 2016) and Lazarian et al. (2022) are not included. The study by Houde et al. (2009) is an attempt to fit the correction factor assuming isotropic turbulence cells and that of Lazarian et al. (2022) depends on velocity gradients. It would be desirable to compare the results in this paper with those in Houde et al. (2009, 2016) and in Lazarian et al. (2022), in a separate publication.

**2.1. "Original" DCF** The version proposed by Davis (1951) and by Chandrasekhar & Fermi (1953; hereafter DCF) can be written $B_{\rm DCF,o} = (4\pi\rho)^{1/2} \sigma_v / \sigma_\theta$ where $\rho$ is the gas density. Here $\sigma_v$ is the turbulent velocity dispersion which excites Alfvén waves on the initial magnetic field lines, assumed to be uniform. The rms wave amplitude perpendicular to the mean field direction $\sigma_{\delta B_\perp}$ is a fraction of the mean field strength $B_0$. This fraction is approximated as $\sigma_\theta$, the standard deviation of field directions in the plane of the sky, in the limit of small-amplitude waves. This pioneering equation does not consider any complicating details of spatial structure in the magnetic field, the density, or the velocity dispersion. Comparison of this equation with MHD turbulent numerical simulations indicates a field strength correction factor ~0.5 for $\sigma_\theta \lesssim 25°$ (OSG01).

**2.2. "Large-angle" DCF** The more general version $B_{\rm DCF,la} = (4\pi\rho)^{1/2} \sigma_v / \sigma_{\tan\theta}$ replaces $\sigma_\theta$ in the original DCF equation with $\sigma_{\tan\theta}$, to relax the DCF assumption of small angles (Heitsch et al. 2001, Liu et al. 2021). However, this version significantly underestimates the mean field when the mean field is much weaker than the amplitude of its fluctuating component. In that case large-angle values of $\tan\theta$ dominate $\sigma_{\tan\theta}$. According to MHD turbulent simulations the mean DCF correction factor is ~0.3 for fields stronger than normalized field strength $B_{\text{model}} = 1.2$, but this factor increases by a factor ≳10 for weaker fields (Heitsch et al. 2001, Figure 6). In the small-angle approximation of this equation, comparison with a strong-field simulation also indicates a mean DCF correction factor 0.3 (Liu et al. 2021 Table 6, Liu et al. 2022).

**2.3. "Standard" DCF** In this version $B_{\text{DCF,s}} = (4\pi\rho)^{1/2}\sigma_v/\tan\sigma_\theta$, the original DCF values of density and velocity dispersion are replaced by their mean values over the map. The term $\sigma_{\tan\theta}$ in the large-angle DCF formula is replaced by $\tan\sigma_\theta$ based on an equipartition argument, providing more realistic correction factors for weak fields (Falceta-Gonçalves et al. 2008, Li et al. 2021, C22). When the number of independent map points is too small to apply the most-likely method, this standard DCF equation is recommended, over the range $0 \leq \sigma_\theta \lesssim 40°$. Then multiplication by a factor 0.5 - 1 is recommended to account for projection from 3D to 2D angles, depending on the smoothness of angle variation along the line of sight (C22 Figure 16).

The standard DCF equation is the preferred DCF equation in the first simulation study to obtain most-likely field strength estimates numerically (C22).

**2.4. "Structure-function" DCF** This version $B_{\text{DCF,sf}} = (4\pi\rho)^{1/2}\sigma_v\left(\sigma_\theta^{-2} - 1\right)^{1/2}$ corrects for relatively simple structure in the ordered component of the plane-of-the-sky field strength, based on a two-point correlation analysis (Hildebrand et al. 2009, Li et al. 2022). This expression for $B_{\text{DCF,sf}}$ is based on equation (7) of Hildebrand et al. (2009) and on equation (A28) of Li et al. (2022). In this expression it is assumed that the DCF equation coefficient is $f_{\text{DCF}} = 1$ as explained earlier in this section. It is also assumed that the turbulent correlation length is small enough so that the angular dispersion function is well fit by the first two terms of the structure function model.

**2.5. "Parallel - $\boldsymbol{\delta B}$" DCF** This version $B_{\text{DCF},\parallel\delta B} = (2\pi\rho)^{1/2}\sigma_v\sigma_\theta^{-1/2}$ is based on the premise that ISM turbulence is anisotropic rather than isotropic, and that non- Alfvénic or compressible magnetic wave modes may be important in generating the observed polarization. The turbulent motions are in approximate equipartition with the parallel component of the perturbed field, in contrast to Alfvénic DCF models. Comparison with MHD simulations indicates that the resulting field strength predictions slightly underestimate the mean field strength. These results indicate correction factors slightly greater than unity, in contrast to Alfvénic wave equations whose correction factors are significantly less than unity (Skalidis & Tassis 2021).

## 3. Probability Distributions of the Local-to-Global Field Strength Ratio

This section derives probability distribution $p(\beta)$, where $\beta = B_{\mathrm{DCF,local}}/B_{\mathrm{DCF}}$ is the ratio of the local and global versions of DCF equations in a map region. This treatment assumes that the density and velocity dispersion each have relatively small variation over the map compared to the polarization angles. The polarization angles follow a standard normal distribution with standard deviation $\sigma_\theta$ in the range 0 - 30°.

The assumed 2D normal distribution of polarization angles in the plane of the sky arises only for restricted conditions, where the 3D angle deviations follow a normal distribution with dispersion $\sigma_\theta$, and where the inclination of the mean 3D field is less than ~50 deg. The 3D vectors must also have a relatively "smooth" rather than random distribution along the line of sight so that their line-of-sight cancellation is negligible (C22). Under these conditions the 2D angle deviations follow essentially the same normal distribution as the 3D angle deviations (C22 Figure 16).

The next subsection derives $p(\beta)$ and $\beta_{\mathrm{ml}}$ for each of the five DCF equations described above in Section 2. It summarizes the properties of the equations and their modal values in Table 1 and in Figures 1 - 5. It compares the derived values of $\beta_{\mathrm{ml}}$ with corresponding values $\beta_{\mathrm{sim}}$ from simulation studies in Table 2.

### 3.1. Probability Distributions and Most-Likely Correction Factors

The ratio $\beta$ of the local to the global DCF field strength estimates in a polarization map can be written as the ratio of the denominators of the global and local DCF equations, $\beta \equiv \delta_{\mathrm{global}}/\delta_{\mathrm{local}}$. Here $\delta_{\mathrm{global}} \equiv (4\pi\rho)^{1/2}\sigma_v/B_{\mathrm{DCF}}$ and $\delta_{\mathrm{local}}$ is its local analog. The most likely value of $\beta$ is its modal value $\beta_{\mathrm{ml}}$ obtained by setting the derivative of $p(\beta)$ to zero. Assuming $B_{\mathrm{ml}}$ is the statistically most accurate implementation of a DCF equation (C22), $B_{\mathrm{ml}} = \beta_{\mathrm{ml}} B_{\mathrm{DCF}}$ where the modal value $\beta_{\mathrm{ml}}$ is the correction factor to be applied to $B_{\mathrm{DCF}}$. C22 applied this concept to **the** numerical evaluation of the "standard" DCF equation described in Section 2.3. Here this application is extended to each of five DCF equations, using an analytic procedure. Table 1 gives expressions for $\delta_{\mathrm{global}}$, $\delta_{\mathrm{local}}$, $\beta$, and $\beta_{\mathrm{ml}}$ for each DCF equation. Table 1 is similar in its summary of formulas to Table 1 in Liu et al. (2022). The expressions in Table 1 are used to derive and plot the probability distribution $p(\beta)$ for each of the DCF equations in Section 3.11-3.15. These expressions are also used to evaluate the most-likely correction factors $\beta_{\rm ml}$ in Section 3 and Table 2.

To derive each probability distribution $p(\beta)$, a Gaussian probability distribution of polarization angles is assumed, and the change-of-variables relation (Papoulis 1984) is used. For polarization angle $\psi$ and local mean value $\bar{\psi}$ the angle deviation is $\theta \equiv |\psi - \bar{\psi}|$ over $0 \leq \theta \leq \pi/2$ where $\bar{\psi} = 0$. Then the distribution $p(\theta)$ is the one-sided standard normal distribution,

$$p(\theta) = \frac{(2/\pi)^{1/2}}{\sigma_\theta \mathrm{erf}[\pi/(8^{1/2}\sigma_\theta)]} \exp\left[-\frac{1}{2}\left(\frac{\theta}{\sigma_\theta}\right)^2\right] \quad . \tag{1}$$

Sections 3.1.1 to 3.1.5 show the probability distributions $p(\beta)$ derived from equation (1) and from the change of variables relation. Each of these sections gives $p(\beta)$ in equation form and in plot form.

**Table 1.** Most-Likely DCF Expressions

| (1) DCF equation | (2) $\delta_{\rm global}$ | (3) $\delta_{\rm local}$ | (4) $\beta$ | (5) $\beta_{\rm ml}$ | (6) $B_{\rm DCF}$ ref. |
|---|---|---|---|---|---|
| original | $\sigma_\theta$ | $\theta$ | $\sigma_\theta/\theta$ | $2^{-1/2}$ | 1,2,3 |
| large-angle | $\sigma_{\tan\theta}$ | $\tan\theta$ | $\sigma_{\tan\theta}/\tan\theta$ | $\frac{\sigma_{\tan\theta}}{\tan[2\sigma_\theta^2\beta_{\rm ml}/\sigma_{\tan\theta}]}$ | 4,5 |
| standard | $\tan\sigma_\theta$ | $\tan\theta$ | $\tan\sigma_\theta/\tan\theta$ | $\frac{\tan\sigma_\theta}{\tan[2\sigma_\theta^2\beta_{\rm ml}/\tan\sigma_\theta]}$ | 6,7,8 |
| structure function | $(\sigma_\theta^{-2}-1)^{-1/2}$ | $(\theta^{-2}-1)^{-1/2}$ | $\left(\frac{\theta^{-2}-1}{\sigma_\theta^{-2}-1}\right)^{1/2}$ | $[2(1-\sigma_\theta^2)]^{-1/2}$ | 9,7 |
| parallel $\delta B$ | $(2\sigma_\theta)^{1/2}$ | $(2\theta)^{1/2}$ | $(\sigma_\theta/\theta)^{1/2}$ | $(3/2)^{-1/4}$ | 10,7 |
| power-law | $\sigma_\theta^q$ | $\theta^q$ | $(\sigma_\theta/\theta)^q$ | $(1+q)^{-q/2}$ | 11 |

**Notes -** Column (1) gives the name of the DCF equation as in Section 2. In Column (2) $\delta_{\rm global} \equiv \sqrt{4\pi\rho}\sigma_v/B_{\rm DCF}$ is the denominator of the global DCF equation. Here $\rho$ is the mean density, $\sigma_v$ is the mean velocity dispersion over the polarization map, and $B_{\rm DCF}$ is the equation for the plane-of-the-sky field strength according to the references in Column (6). The angle $\theta$ is the deviation of the polarization angle from the mean angle over the map and $\sigma_\theta$ is the standard deviation of $\theta$. In Column (3), $\delta_{\rm local} \equiv \sqrt{4\pi\rho}\sigma_v/B_{\rm local}$ is the local analog of $\delta_{\rm global}$ at each point in the polarization map. Here the standard deviation is replaced by the local deviation, as in C22. Column (4) gives the ratio $\beta$ of the local and global estimates of $B_{\rm DCF}$, $\beta = \delta_{\rm global}/\delta_{\rm local}$. Column (5) gives the DCF correction factor $\beta_{\rm ml}$, the most-likely value of $\beta$, based on the mode of the distribution $p(\beta)$ given in the text. The large-angle and standard expressions are to be solved for $\beta_{\rm ml}$ numerically. They each approach $\beta_{\rm ml} = 2^{-1/2}$ in the limit of small $\sigma_\theta$. The structure function expression for $\beta_{\rm ml}$ is an approximation based on the assumption $\sigma_\theta \ll 3^{-1/2}$. The power-law expression applies to the original DCF equation when $q = 1$ and to the parallel $\delta B$ equation when $q = 1/2$, as described in Section 3.2. It applies approximately to the large-angle, standard, and structure function equations in the small-angle limit when $q = 1$. References - (1) Davis (1951); (2) Chandrasekhar & Fermi (1953); (3) Ostriker et al. (2001); (4) Heitsch et al. (2001), (5) Liu et al. (2021); (6) Falceta-Gonçalves et al. (2008); (7) Li et al. (2021); (8) Chen et al. (2022 (C22)); (9) Hildebrand et al. (2009); (10) Skalidis & Tassis (2021); (11) Section 3.2 of this work.

### 3.1.1. $p(\beta)$ for the Original DCF Equation

For the original DCF equation where $\beta = \sigma_\theta/\theta$ (Section 2.1 and Table 1), and for $p(\beta) = |d\theta/d\beta|\, p(\theta)$ one obtains the probability distribution of the local to global field strength ratio

$$p(\beta) = c_\beta \frac{\exp\left(-\frac{1}{2\beta^2}\right)}{\beta^2} \quad (2)$$

where the normalizing constant is $c_\beta \equiv (2/\pi)^{1/2}/\mathrm{erf}[\pi/(8^{1/2}\sigma_\theta)]$. For values of angle dispersion $\sigma_\theta$ considered here, the error function (erf) is negligibly different from unity, so equation (2) is essentially independent of $\sigma_\theta$. This probability distribution is shown in Figure 1. It matches the positive portion of the reciprocal normal distribution (Johnson et al. 1994).

The "most-likely" value of $\beta$ in equation (2) occurs at the modal value $\beta_{\rm ml} = 2^{-1/2} = 0.71$, which is obtained by setting the derivative $dp/d\beta$ of equation (2) to zero.

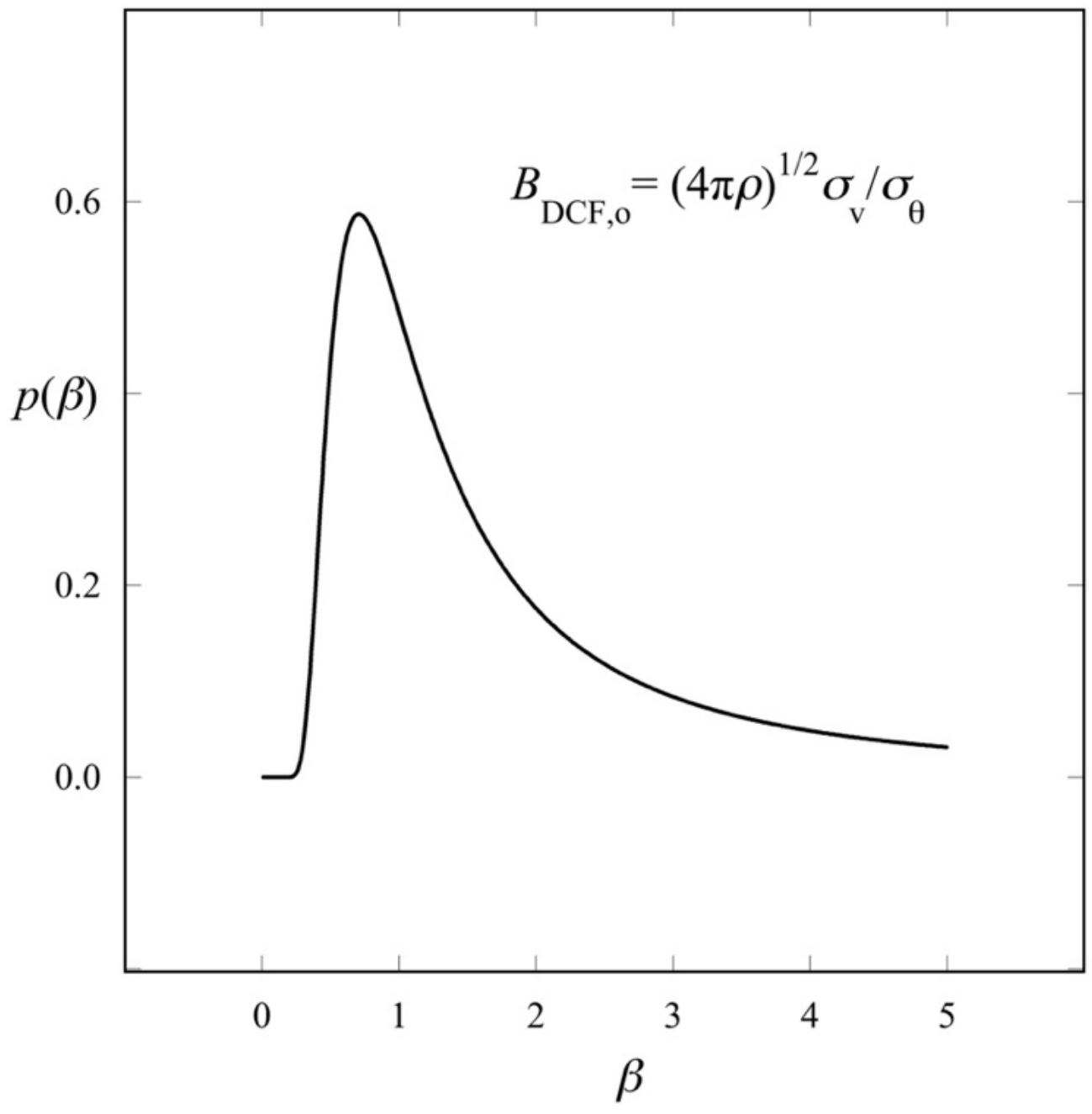

**Figure 1.** Probability distribution $p(\beta)$ for the ratio $\beta$ of local to global plane-of-the-sky field strength, for the "original" DCF equation (Davis 1951; Chandrasekhar & Fermi 1953; OSG01) denoted here as $B_{\rm DCF,o}$. This distribution has modal value $\beta_{\rm ml} = 2^{-1/2} = 0.71$, independent of angle dispersion $\sigma_\theta$.

### 3.1.2. $p(\beta)$ for the Large-Angle DCF Equation

For the large-angle DCF equation (Section 2.1) the change of variables relation in Section 3.1.1 gives

$$p(\beta) = c_\beta \frac{\sigma_{\tan\theta}}{\sigma_\theta} \frac{\exp\left\{-\frac{1}{2\sigma_\theta^2}\left[\tan^{-1}\left(\frac{\sigma_{\tan\theta}}{\beta}\right)\right]^2\right\}}{\beta^2+\sigma_{\tan\theta}^2} \quad . \qquad (3)$$

This function is shown in Figure 2 for four values of $\sigma_\theta$. Differentiation gives the equation for the modal value, summarized in Table 1. Numerical values of $\beta_{\rm ml}$ are close to 0.7, as given in Table 2, for the small range of $\sigma_\theta$ up to ~15°. At larger values of $\sigma_\theta$, the distribution shape changes dramatically and the tail of the distribution becomes dominant. Then $\beta_{\rm ml}$ and $B_{\rm ml}$ become unrealistically large, as is also seen in simulations (Heitsch et al. 2001). This behavior occurs because as $\sigma_\theta$ increases, high values of $\tan\theta$ dominate $\sigma_{\tan\theta}$ (Heitsch et al. 2001, Falceta-Gonçalves et al. 2008, Li et al. 2022).

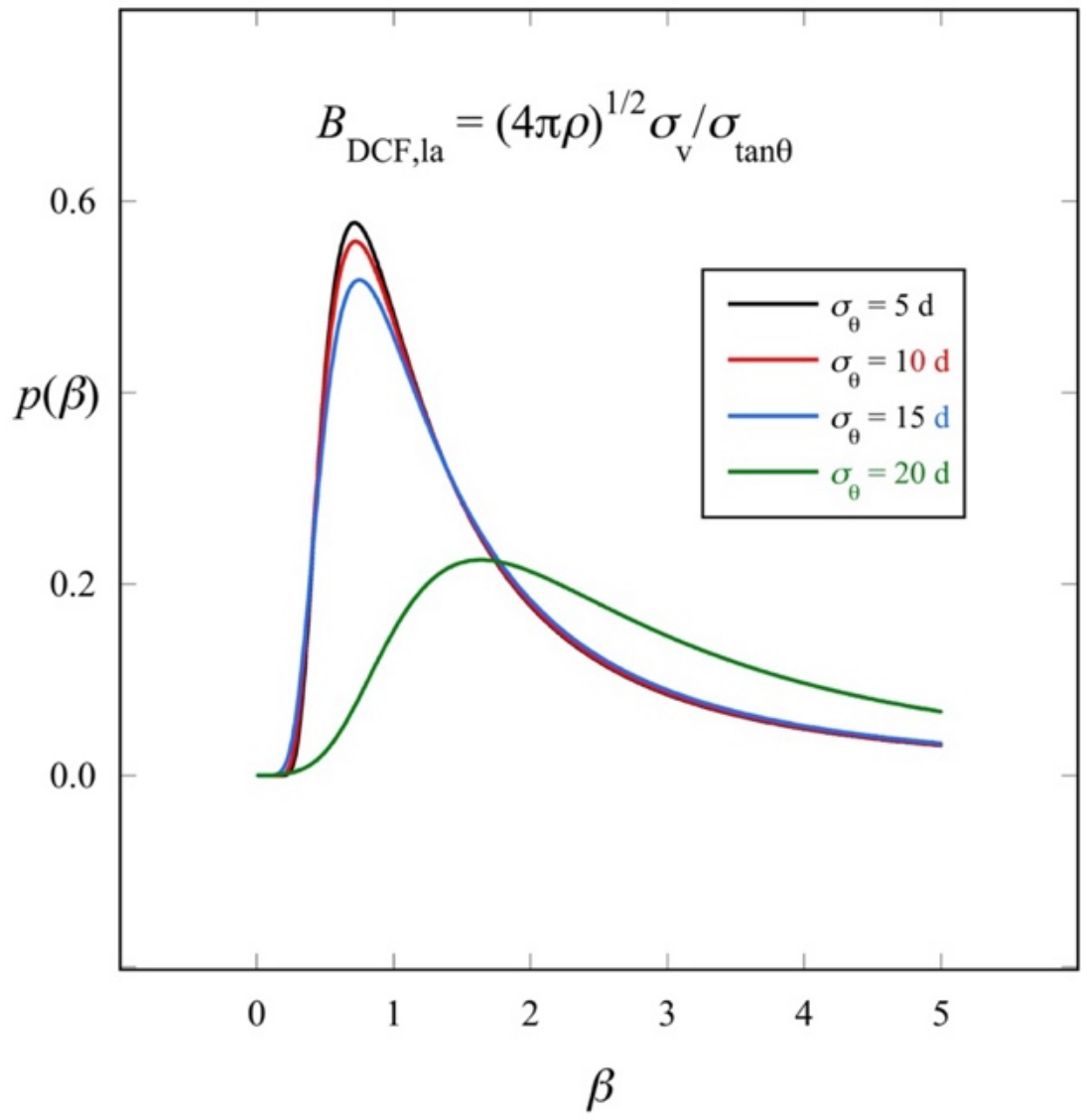

**Figure 2.** Probability distributions $p(\beta)$ for the ratio $\beta$ of local to global plane-of-the-sky field strength, for the "large-angle" DCF equation (Heitsch et al. 2001, Liu et al. 2021). These distributions have similar shape, and similar modal values close to 0.7 only for relatively small angle dispersions $\sigma_\theta = 5$ deg *(black)*, 10 deg *(red)*, and 15 deg *(blue)*. For larger $\sigma_\theta$ values, e.g. 20 deg *(green)*, the distribution broadens and the modal value increases rapidly.

### 3.1.3. $p(\beta)$ for the Standard DCF Equation

This equation has nearly the same form as equation (3), with $\tan\sigma_\theta$ replacing $\sigma_{\tan\theta}$. In contrast to Figure 2, its curves in Figure 3 retain a similar shape over a wider range of angle dispersions $\sigma_\theta$,

$$p(\beta) = c_\beta \frac{\tan\sigma_\theta}{\sigma_\theta} \frac{\exp\left\{-\frac{1}{2\sigma_\theta^2}\left[\tan^{-1}\left(\frac{\tan\sigma_\theta}{\beta}\right)\right]^2\right\}}{\beta^2+(\tan\sigma_\theta)^2} \quad . \qquad (4)$$

Differentiation gives its modal equation, summarized in Table 1, with numerical values in Table 2.

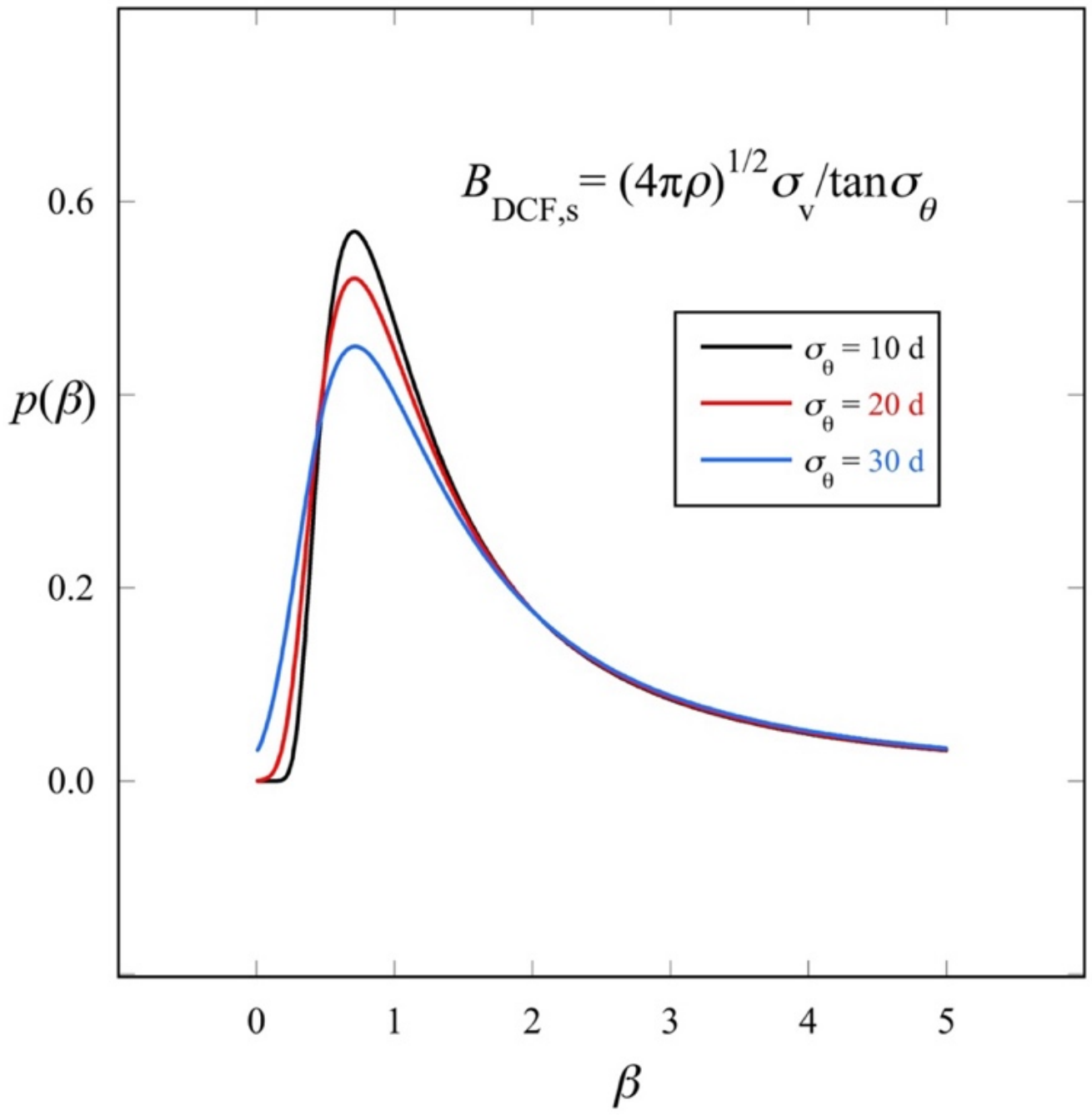

**Figure 3.** Probability distributions $p(\beta)$ for the ratio $\beta$ of local to global plane-of-the-sky field strength, for the "standard" DCF equation described by Falceta-Gonçalves et al. (2008), Li et al. (2022), and C22. These distributions vary slightly as angle dispersion $\sigma_\theta$ increases from 10 degrees (*black*) to 20 degrees (*red*) to 30 degrees (*blue)*, with modal values close to 0.7.

### 3.1.4. $p(\beta)$ for the Structure Function DCF Equation

This equation has the form

$$p(\beta) = c_\beta \frac{\beta(1-\sigma_\theta^2)}{[\sigma_\theta^2+\beta^2(1-\sigma_\theta^2)]^{3/2}} \exp\left\{-\frac{1}{2[\sigma_\theta^2+\beta^2(1-\sigma_\theta^2)]}\right\} \quad . \qquad (5)$$

Its curves in Figure 4 have similar shape for $\sigma_\theta = 10° - 30°$, with similar modes to those in Figures (1), (2), and (3). Its modal equation is given in Table 1, with numerical values in Table 2. Equations (3), (4), and (5) all reduce to the reciprocal normal equation (2) in the small-angle limit $\sigma_\theta \ll 1$.

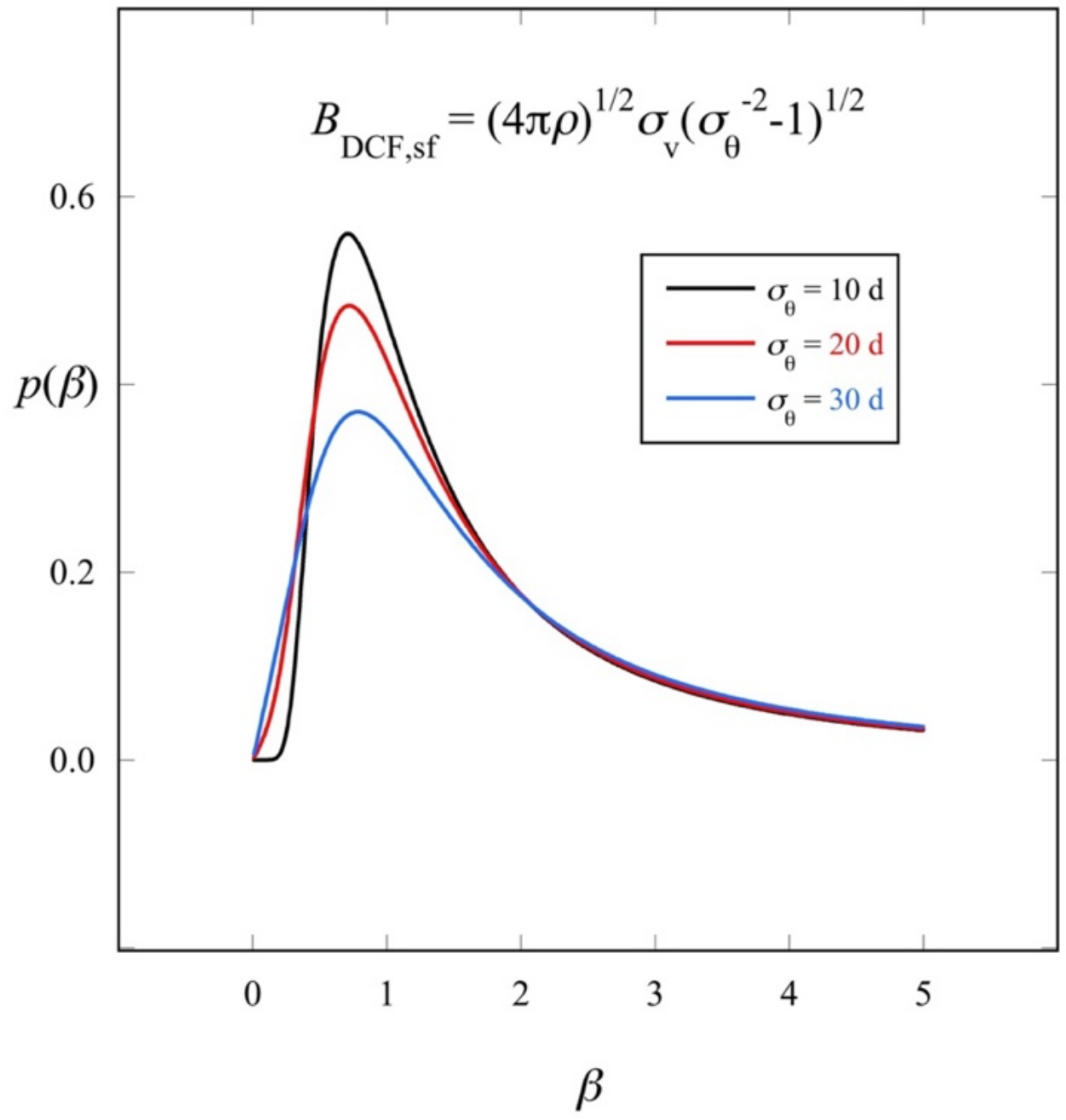

**Figure 4.** Probability distributions $p(\beta)$ for the ratio $\beta$ of local to global plane-of-the-sky field strength, for the structure-function DCF equation of Hildebrand et al. (2009) and Li et al. (2022). These distributions vary slightly as angle dispersion $\sigma_\theta$ increases from 10 degrees (*black*) to 20 degrees (*red*) to 30 degrees (*blue)*, with modal values close to 0.7.

### 3.1.5. $p(\beta)$ for the Parallel-$\delta B$ DCF Equation

This equation differs from equations (2) - (5) because it is based on non-Alfvénic MHD modes driven by turbulent motions parallel to the mean field direction (Skalidis & Tassis 2021). It depends on $\sigma_\theta$ as $\sigma_\theta^{-1/2}$ rather than as $\sigma_\theta^{-1}$ for the Alfvénic equations (2)-(5). The resulting probability distribution is

$$p(\beta) = c_\beta \frac{2\exp\left(\frac{-1}{2\beta^4}\right)}{\beta^3} \quad . \qquad (6)$$

The $p(\beta)$ curve in Figure (6) has a narrower shape and a slightly larger modal value than do the curves for DCF equations based on turbulent driving of Alfvénic MHD waves in Figures (1) - (4). The modal value $\beta_{\rm ml} = (3/2)^{-1/4} = 0.90$ is listed in Tables 1 and 2. The form of this distribution is less well-known than that of equation (2). It is neither a special case of the reciprocal normal distribution nor of the generalized inverse normal distribution (Johnson et al. 1994).

As for the original DCF equation (Sections 2.1, 3.1.1), equation (6) is essentially independent of $\sigma_\theta$ except for a weak dependence in the normalizing factor $c_\beta$.

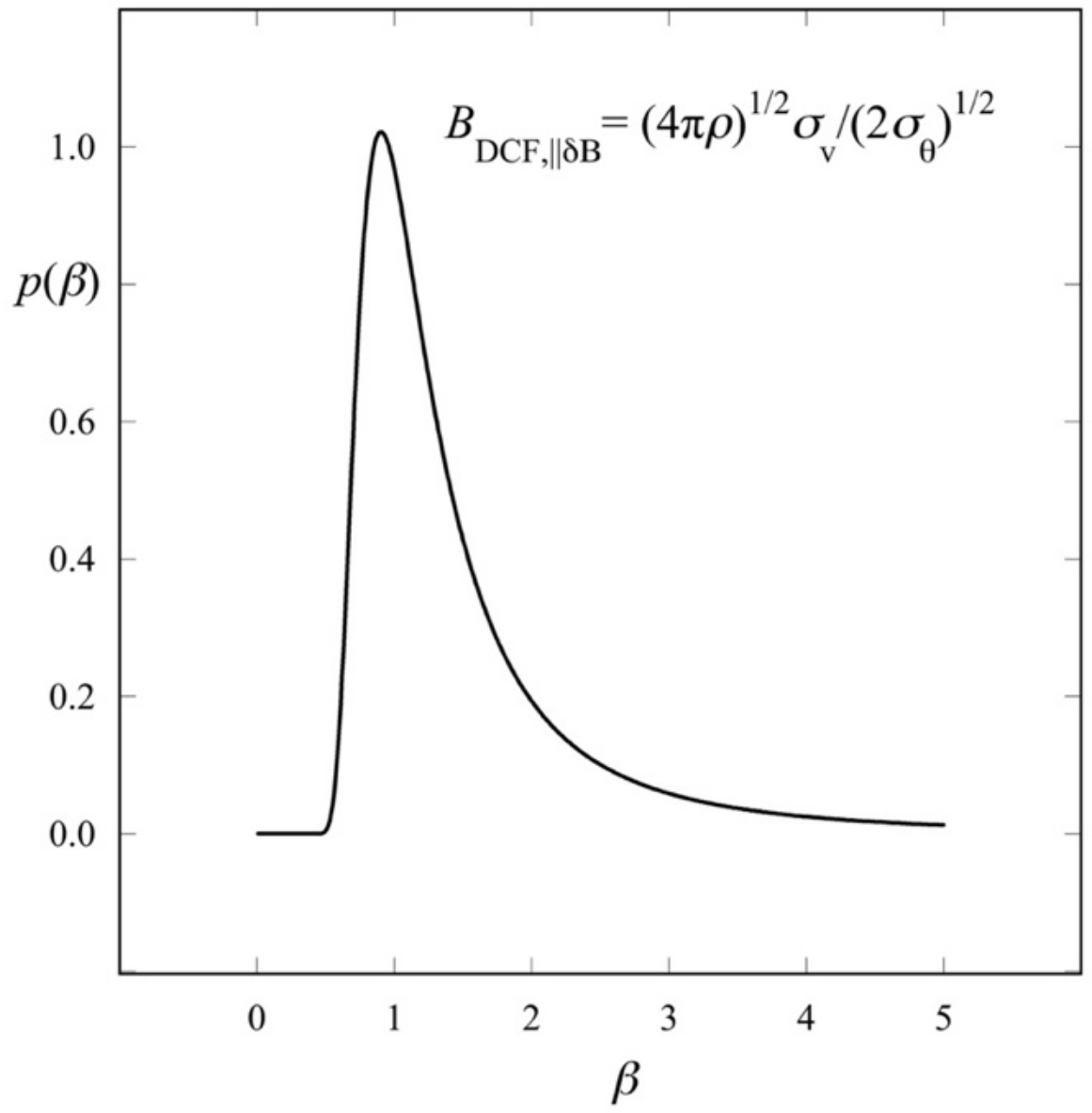

**Figure 5.** Probability distribution $p(\beta)$ for the ratio $\beta$ of local to global plane-of-the-sky field strength, for the "parallel-$\delta B$" DCF. This equation assumes that non-Alfvénic MHD modes are driven by turbulent motions parallel to the mean field. The field strength depends on the angle dispersion as $\sigma_\theta^{-1/2}$ rather than as $\sigma_\theta^{-1}$ as in the original DCF equation (Skalidis & Tassis 2021). The resulting distribution is narrower, and its modal value $\beta_{\rm ml} = (3/2)^{-1/4} = 0.9$ is slightly greater than in Figures 1-4 for DCF equations which represent excitation of Alfvénic modes.

### 3.2. Interpretation of Most-Likely Distributions and Modes

The $p(\beta)$ distributions of Section 3.1 have similar shapes and a narrow range of modes, from $\beta_{\rm ml} \approx 0.7$ to 0.9. Their shapes are similar because each distribution is the product of a declining exponential and a declining power law. The shapes of the Alfvénic curves vary slightly since each Alfvénic DCF equation has slightly different dependence on angle dispersion $\sigma_\theta$. As $\sigma_\theta$ decreases, all of the Alfvénic curves approach the same shape.

The similarity of modal values from one DCF equation to the next can be understood from the power-law form of the DCF equations, $B_{\mathrm{DCF,pl}} = (4\pi\rho)^{1/2}\sigma_v/\sigma_\theta^q$. When $q = 1$ this relation matches the form of the original DCF equation and of the other Alfvénic equations in the small-angle limit. When $q = 1/2$ it matches the form of the parallel-$\delta B$ equation. Then $\beta$ can be written $\beta = (\sigma_\theta/\theta)^q$ so that $p(\beta)$ is obtained from the change of variables relation as in equations (2) - (6). Solving $dp(\beta)/d\beta = 0$ yields a simple expression for the most-likely value of $\beta$,

$$\beta_{\mathrm{ml}} = (1+q)^{-q/2} \quad . \qquad (7)$$

In equation (7), substitution of $q = 1$ or $q = 1/2$ yields the values derived above, $\beta_{\mathrm{ml}} = 2^{-1/2}$ or $\beta_{\mathrm{ml}} = (3/2)^{-1/4}$.

Equation (7) shows that $\beta_{\mathrm{ml}}$ increases toward a maximum value of 1 as $q \to 0$. This property is exemplified by the increase from $\beta_{\mathrm{ml}} = 0.71$ when $q = 1$ to $\beta_{\mathrm{ml}} = 0.90$ when $q = 1/2$. This property also implies that the DCF equation for a cloud with uniform density, uniform velocity dispersion, and a normal angle distribution cannot have a most-likely correction factor exceeding $\beta_{\mathrm{ml}} = (1+q)^{-q/2}$.

On the other hand, $\beta_{\mathrm{ml}}$ may in principle exceed $(1+q)^{-q/2}$ if the density is correlated with the angle dispersion. For example in a self-gravitating region which forms by flux freezing and mass conservation, the density becomes centrally concentrated and the field lines pinch inward. Then the angle dispersion over the map area increases with increasing central density (Mestel 1966). In this case the density in the numerator of the DCF equation is correlated with the angle dispersion in the denominator. This correlation reduces the net dependence of the DCF equation on angle dispersion. If so, the effective value of $q$ decreases, allowing $\beta_{\mathrm{ml}}$ to increase above its limit for uniform density and velocity dispersion.

The most-likely correction factors derived here are based on a probabilistic treatment of the angle dependence of the DCF equation. They are independent of the physical effects which have been suggested to interpret the values of correction factors determined by MHD simulations, described in Section 1. These statistical and physical correction factors are compared in Table 2 since they all lie within the range 0.3 - 1.2, and they all appear relevant for DCF estimation of field strengths from observations.

### 3.3 Comparison of Correction Factors

In this section, the foregoing expressions for $\beta_{\mathrm{ml}}$ in Table 1 are evaluated in numerical form over the range of angle deviations where each DCF equation is valid, over $0 \leq \sigma_\theta \leq 30°$ as for the original DCF equation (OSG01) or over $0 \leq \sigma_\theta \leq 15°$ for the large-angle equation (Heitsch et al. 2001). The values of $\beta_{\mathrm{sim}}$ are drawn from five turbulent MHD simulation studies with a total of 17 simulation runs. For each DCF equation, Table 2 presents estimates of the difference between mean values of $\beta_{\mathrm{ml}}$ and $\beta_{\mathrm{sim}}$. Assuming that $\beta_{\mathrm{sim}}$ is set by independent statistical and physical causes, Table 2 also presents an estimate $\gamma_{\mathrm{ml}}$ of the relative field strength correction due to statistical causes.

The selection of simulation correction factors in Table 2 is more extensive than that in Table 3 in Liu et al. (2022). Table 2 includes correction factors from three simulation runs by Heitsch et al. (2001), two by Li et al. (2022), six by C22, and five by Skalidis & Tassis (2021) which were not reported in Liu et al. (2022), and one from OSG01, which was reported by Liu et al. (2022). Each simulation is a 3D calculation which includes signal integration along the line of sight.

Each simulation study reporting $\beta_{\mathrm{sim}}$ for a particular DCF equation is represented in Table 2 by a single reported value of $\beta_{\mathrm{sim}}$ (references 3,7) or by an average value over multiple simulation runs (references 4, 5, 8, and 10). The average over multiple runs is intended to obtain the most representative value. Selection of values of $\beta_{\mathrm{sim}}$ from within a group of multiple runs is discussed in Section 4.1. In the largest group of multiple runs, six values of $\beta_{\mathrm{sim}}$ are chosen from reference (8) based on their "eq" method of density estimation with a model of self-gravitating cloud equilibrium. This method is chosen because it is less sensitive to choice of DCF equation than the "vfit" method based on a power-law relation between spectral line velocity and cloud map radius (C22).

Table 2 reports the difference $\Delta\beta$ between the mean values of $\beta_{\mathrm{ml}}$ and $\beta_{\mathrm{sim}}$ for each DCF equation. If **each** value of $\beta_{\mathrm{sim}}$ for a given DCF equation were due entirely to the statistical most-likely property, i.e. if $\beta_{\mathrm{ml}} = \beta_{\mathrm{sim}}$, $\Delta\beta$ would equal zero for that DCF equation. In Table 2 Column (4) the values of $\Delta\beta$ range from -0.3 to 0.4. Their mean over all five equations is $\overline{\Delta\beta} = 0.08$ with standard error 0.12.

As a check, the mean over all values of $\Delta\beta$ was calculated by assigning equal weight to the values of $\beta_{\mathrm{ml}}$ and of $\beta_{\mathrm{sim}}$ associated with each of the 17 simulation runs. Then using the same procedure as in Table 2, $\overline{\Delta\beta}$ $=-0.01 \pm 0.07$. As a second check, substitution of "vfit" density estimates for "eq" estimates in C22 yields $\overline{\Delta\beta}$= 0.05 ± 0.13 for equal weighting by DCF equation and $\overline{\Delta\beta}$= -0.13 ± 0.09 for equal weighting by simulation. Thus $\beta_{\mathrm{sim}}$ and $\beta_{\mathrm{ml}}$ appear consistent on average over all equations or over all simulation runs, but they disagree significantly for the large-angle and parallel-$\delta B$ DCF equations.

This pattern of partial agreement is also shown by the ranges of $\beta_{\mathrm{ml}}$ and $\beta_{\mathrm{sim}}$. Table 2 shows that among the five DCF equations, the ratio of maximum to minimum values for $\beta_{\mathrm{ml}}$ is ~1.3. This range is set by the similarity of the dependence of $B_{\mathrm{DCF}}$ on $\sigma_\theta$ among the five DCF equations, as noted in Section 3.2. In contrast, the values of $\beta_{\mathrm{sim}}$ have a range of a factor of ~4. This greater range may be due in part to the variety of initial conditions and parameter values among the simulations selected for Table 2. The dependence of $\beta_{\mathrm{sim}}$ on simulation properties is discussed in more detail by Liu et al. (2022).

The foregoing results indicate that $\beta_{\mathrm{ml}}$ alone is not an accurate predictor of $\beta_{\mathrm{sim}}$.. Thus it appears more realistic to treat the correction from $B_{\mathrm{DCF}}$ to $B_{\mathrm{sim}}$ as a combination of a statistical component and a non-statistical component, which has a physical origin. Therefore in Table 2, Column (5) reports the "relative statistical correction" $\gamma_{\mathrm{ml}}$, a measure of the relative importance of the statistical component of the correction. Here "correction" refers to a difference in field strengths, in contrast to "correction factor" which refers to a ratio of field strengths.

The overall correction from $B_{\mathrm{DCF}}$ to $B_{\mathrm{sim}}$ is assumed to consist of two parts, a correction $B_{\mathrm{DCF}} - B_{\mathrm{ml}}$ of statistical origin, and a correction $|B_{\mathrm{ml}} - B_{\mathrm{sim}}|$ of physical origin. Here the absolute value notation is needed to keep each correction positive, since $B_{\mathrm{sim}} > B_{\mathrm{ml}}$ for the non-Alfvénic DCF equation. Then $\gamma_{\mathrm{ml}}$ is written in terms of the corrections given above as

$$\gamma_{\mathrm{ml}} = \frac{B_{\mathrm{DCF}} - B_{\mathrm{ml}}}{B_{\mathrm{DCF}} - B_{\mathrm{ml}} + |B_{\mathrm{ml}} - B_{\mathrm{sim}}|} \tag{8}$$

or in terms of the corresponding correction factors as

$$\gamma_{\mathrm{ml}} = \left(1 + \left|\frac{1-\beta_{\mathrm{sim}}}{1-\beta_{\mathrm{ml}}} - 1\right|\right)^{-1} \quad . \tag{9}$$

In equations (8) and (9) the statistical correction dominates the physical correction when $\gamma_{\mathrm{ml}} > 1/2$. The relative physical correction $\gamma_{\mathrm{phys}}$ corresponding to $\gamma_{\mathrm{ml}}$ is $\gamma_{\mathrm{phys}} = 1 - \gamma_{\mathrm{ml}}$. Note that $\gamma_{\mathrm{ml}}$ indicates the relative importance of the statistical and physical corrections, but it is not a simple ratio of the correction factors $\beta_{\mathrm{ml}}$ and $\beta_{\mathrm{sim}}$.

Column (5) of Table 2 shows that the statistical correction is a significant fraction of the total correction for all five DCF equations. It ranges from $\gamma_{\mathrm{ml}} = 0.3$ for the parallel-$\delta B$ equation to $\gamma_{\mathrm{ml}} = 0.9$ for the standard DCF equation. The mean and median values of $\gamma_{\mathrm{ml}}$ are each $\gamma_{\mathrm{ml}} = 0.6$. The greatest values of $\gamma_{\mathrm{ml}}$ are 0.6 for the original DCF equation, 0.7 for the structure function equation and 0.9 for the standard DCF equation. For these equations, most of the typical simulation correction is matched by the statistical correction, with relatively less physical correction required. This result is consistent with the small values of $\Delta\beta \leq 0.2$ in Column (4) of Table 2.

**Table 2.** Most-Likely and Simulation Correction Factors for Five DCF Equations

| (1) DCF equation | (2) $\beta_{\mathrm{ml}}$ $\sigma_{\theta 1}$ $\sigma_{\theta 2}$ $\sigma_{\theta 3}$ | (3) $\beta_{\mathrm{sim}}$ | (4) $\Delta\beta$ | (5) $\gamma_{\mathrm{ml}}$ | (6) sim. ref. |
|---|---|---|---|---|---|
| original | 0.71 | 0.49 | 0.2 | 0.6 | 3 |
| large-angle | 0.71 0.72 0.75 | 0.33, 0.28 | 0.4 | 0.4 | 4,5 |
| standard | 0.69 0.71 0.72 | 0.71, 0.74 | 0.0 | 0.9 | 7,8 |
| structure function | 0.72 0.75 0.83 | 0.66 | 0.1 | 0.7 | 7 |
| parallel $\delta B$ | 0.90 | 1.18 | -0.3 | 0.3 | 10 |

**Notes** - Column (1) gives the name of the DCF equation as described in Section 2. Column (2) gives values of the most-likely correction factors $\beta_{\mathrm{ml}}$ for angle standard deviations $\sigma_{\theta 1}$, $\sigma_{\theta 2}$, and $\sigma_{\theta 3}$, which span the range of allowed values. Within Column (2), when $\beta_{\mathrm{ml}}$ is independent of $\sigma_\theta$, Rows (1) and (5) each give a single value of $\beta_{\mathrm{ml}}$. When $\beta_{\mathrm{ml}}$ depends on $\sigma_\theta$, Row 2 gives $\beta_{\mathrm{ml}}$ for angles $\sigma_\theta = 5°$, $10°$, and $15°$, while Rows (3) and (4) give $\beta_{\mathrm{ml}}$ for $\sigma_\theta = 10°$, $20°$, and $30°$. Column (3) gives correction factors $\beta_{\mathrm{sim}}$ according to the simulations referenced in Column (5). When the reference reports multiple simulation values, Column (3) gives the mean value. When there are two references for one DCF equation, Column (3) gives the two values of $\beta_{\rm sim}$ separately. Column (4) gives for each DCF equation the correction factor difference between the mean of the $\beta_{\rm ml}$ values in Column (2) and the mean of the $\beta_{\rm sim}$ values in Column 3, $\Delta\beta \equiv \overline{\beta_{\rm ml}} - \overline{\beta_{\rm sim}}$. Column (5) gives $\gamma_{\rm ml}$, the relative statistical correction, computed from equation (9). Column (6) gives simulation references using the same numbers as in Table 1.

Additional simulation information: for reference (3) from OSG01, angle dispersions are restricted to $\sigma_\theta \lesssim 25°$. For (4) from Heitsch et al. (2001) Figure 6, $\beta_{\rm sim}$ ranges from 0.3 to 0.4, with mean value 0.3, for fields stronger than the normalized field strength $B_{\rm model} = 1.2$. For (7) from Li et al. (2021), $\beta_{\rm sim}$ is the ratio $B_{\rm 0,true}/B_{\rm 0,DCF}$ from their Table 3, columns "DCF" and "DCF/SF", assuming $f_{\rm DCF} = 1$ as explained in Section 2 of this paper. For (8) from C22, $\beta_{\rm sim}$ ranges from 0.5 to 1.0, with mean value 0.7, based on six estimates in their Table 2 . They represent three simulations with $\sigma_\theta < 30°$, inclination <60°, with correction for 3D to 2D projection, for the density estimation method "eq." Assuming that the cloud is a self-gravitating layer in hydrostatic equilibrium, this "eq" method yields a correction factor $B_{\rm DCF,ml}/\bar{B}$ ranging from ~0.5 to 1.0. Assuming instead that the cloud follows a power-law relation between line velocity and map radius, the "vfit" method yields a typical correction factor ranging from ~0.8 to 1.5. For (10) from Skalidis & Tassis (2021), where $\beta_{\rm sim}$ ranges from 0.91 to 1.38, with mean value 1.18, based on $V_{\rm A}^{\rm true}/V_{\rm A}^{\rm new}$ for the five simulations listed in their Table 1.

## 4. Discussion

### 4.1. Uncertainties in $\boldsymbol{\beta}_{\rm ml}$, $\Delta\boldsymbol{\beta}$, and $\boldsymbol{\gamma}_{\rm ml}$

The expressions for $\beta_{\rm ml}$ in Section 2 assume that the angle deviations $\theta$ and standard deviations $\sigma_\theta$ are known exactly, but in typical observations they are subject to statistical and measurement error. This section presents the relative uncertainty in $\beta_{\rm ml}$ for any power-law DCF equation where $\beta = (\sigma_\theta/\theta)^q$ as given in Table 1 and where $\beta_{\rm ml} = (1+q)^{-q/2}$ in equation (7). The measurement uncertainty in $\theta$ is denoted $\delta\theta$ and the measurement uncertainty in $\sigma_\theta$ is denoted $\delta\sigma_\theta$. It is assumed that the relative measurement uncertainty in $\sigma_\theta$ is significantly less than the relative measurement uncertainty in $\theta$, i.e. $\delta\sigma_\theta/\sigma_\theta \ll \delta\theta/\theta$. Then propagation of errors gives the uncertainty in $\beta$ as $\delta\beta \approx |\partial\beta/\partial\theta|\delta\theta$. The resulting uncertainty in $\beta$ at its modal value equals the product of a function of $q$ and the ratio of the measurement uncertainty $\delta\theta$ to the angle standard deviation $\sigma_\theta$:

$$\delta\beta_{\rm ml} = g(q)\frac{\delta\theta}{\sigma_\theta} \tag{10}$$

where

$$g(q) \equiv q(1+q)^{-(1+q)/2} \ . \tag{11}$$

For the original DCF equation where $q = 1$, $g(q) = 1/2 = 0.50$. For the parallel-$\delta B$ equation where $q = 1/2$, $g(q) = (1/2)(3/2)^{-3/4} = 0.37$. Consequently for the original DCF equation,

$$\beta_{\rm ml,o} = 0.71 \pm 0.50(\delta\theta/\sigma_\theta) \tag{12}$$

and for the parallel-$\delta B$ equation

$$\beta_{\rm ml,par\ \delta B} = 0.90 \pm 0.37(\delta\theta/\sigma_\theta) \ \ . \tag{13}$$

For example when $\delta\theta = 5°$ and $\sigma_\theta = 20°$, $\beta_{\rm ml,o} = 0.7 \pm 0.1$ and $\beta_{\rm ml,par\ \delta B} = 0.9 \pm 0.1$.

The Alfvénic equations in Sections 2.2 - 2.5 are similar to the original equation in Section 2.1 for relatively small angle dispersions. Therefore $g(q) \approx 1/2$ may be sufficiently accurate for error estimation for these equations as well as for the original equation.

Further uncertainties in $\Delta\beta$ and $\gamma_{\rm ml}$ in Table 2 can be estimated from the standard error in the mean for $\bar{\beta}_{\rm ml}$ and $\bar{\beta}_{\rm sim}$. For each mean based on at least three values of $\beta_{\rm ml}$ or $\beta_{\rm sim}$, the typical uncertainties from propagation of errors are $\sigma_{\Delta\beta} = 0.09$ and $\sigma_{\gamma_{\rm ml}} = 0.1 - 0.2$. These uncertainties are useful for distinguishing DCF equations in Section 4.3.

Values of $\bar{\beta}_{\rm sim}$ and $\gamma_{\rm ml}$ also have uncertainty due to selection of simulation runs. For the standard DCF equation, the values of $\beta_{\rm sim} = 0.74$, $\Delta\beta = 0.0$, and $\gamma_{\rm ml} = 0.9$ in Table 2 are based on selection of the six C22 simulation runs having angle dispersion < 30° and mean field inclination < 60° and using the "eq" method of density estimation. This method was selected because it is less sensitive to varying the angle dependence in the DCF equation than is the alternate "vfit" method (C22). Nonetheless if one used the C22 "vfit" results the values in Table 2 would change to $\beta_{\rm sim} = 1.1$, $\bar{\beta}_{\rm sim} = 0.8$, $\Delta\beta = 0.2$, and $\gamma_{\rm ml} = 0.6$ as noted below Table 2. In that case the standard DCF equation would remain in the same group of three equations where the statistical correction is dominant, but it would no longer have the highest rank. Similarly if one selected the minimum value of $\beta_{\rm sim}$ among the five simulation runs in Table 1 of Skalidis & Tassis (2021) rather than the mean value, $\Delta\beta$ would change from -0.3 to -0.01, $\gamma_{\rm ml}$ would change from 0.3 to 0.9, and the parallel $\delta B$ equation would move to the top rank of the group of dominant statistical correction equations.

It is possible that averaging $\beta_{\rm sim}$ over multiple simulation runs in Table 2 may be misleading because the runs are too diverse in their initial conditions and parameters. Instead it may be more informative to identify properties of individual runs whose values of $\beta_{\rm sim}$ agree most closely with $\beta_{\rm ml}$. Among the six runs of C22 described above for the most-likely version of the standard DCF equation (using $\tan\theta$ instead of $\tan\sigma_\theta$), the associated value of $\beta_{\rm ml}$ is $\beta_{\rm ml} = 0.7$ (Table 2, Column 2). The runs with $0.6 \le \beta_{\rm sim} \le 0.8$ with "vfit" or "eq" density estimates do not differ substantially from those with greater or lesser values of $\beta_{\rm sim}$ in angle dispersions or mean densities. However they have distinctly lower values of inclination angle, $\gamma \lesssim 30°$, and velocity dispersion $\sigma_{\rm v} \lesssim 0.3$ km s$^{-1}$. Thus plane-of-the-sky fields and low turbulence may favor agreement between $\beta_{\rm ml}$ and $\beta_{\rm sim}$ for the C22 simulations.

A similar examination of the five simulation runs in Table 1 of Skalidis & Tassis (2021) indicates a significant deviation between mean correction factors for the parallel-$\delta B$ DCF equation, since then $\overline{\beta_{\rm sim}} = 1.2$ while $\overline{\beta_{\rm ml}} = 0.9$. However the best match to $\beta_{\rm ml} = 0.9$ is $\beta_{\rm sim} = 0.91$ as noted above. This match occurs for the run with the lowest sonic Mach number, $M_S = 0.7$, when the corresponding velocity dispersion is $\sigma_{\rm v} = 0.5$ km s$^{-1}$. In contrast the remaining four runs have $\sigma_{\rm v}$ increasing up to 4 km s$^{-1}$. Thus for the parallel-$\delta B$ DCF equation, close agreement between $\beta_{\rm sim}$ and $\beta_{\rm ml}$ occurs for simulation runs with relatively low turbulence, as also found above for the standard DCF equation.

These individual-case comparisons suggest better agreement between $\beta_{\rm sim}$ and $\beta_{\rm ml}$ when simulation physical conditions are similar to those inherent in the most-likely model. These are field directions close to the plane of the sky, and relatively small fluctuations in velocity dispersion, which may be associated with small fluctuations in both density and velocity dispersion. It would be useful to carry out simulations which test this apparent association in more detail.

### 4.2. Limitations

The comparison of $\beta_{\rm ml}$ and $\beta_{\rm sim}$ in Section 3 is limited by a narrow range of $\beta_{\rm ml}$, by a broad range of $\beta_{\rm sim}$, and by their small sample sizes. The presently available simulation studies span a significant range of cloud size scales, initial densities, physical properties, star formation, and methods of generating turbulence. For example in Table 1, simulation reference (1) covers a large region extending to ~8 pc with low densities ~100 $\rm cm^{-2}$ with Alfvén Mach number $M_A = 0.7$ but reference (3) has regions smaller than ~1 pc with much higher densities ~$10^5$ $\rm cm^{-2}$ and a range of $M_A = 0.7 - 2.5$. The simulations with multiple runs have substantial scatter in $\beta_{\rm sim}$ from one run to the next. The distribution of values of $\beta_{\rm ml}$ is not continuous enough to test for correlation between $\beta_{\rm ml}$ and $\beta_{\rm sim}$. Thus the comparison of $\beta_{\rm ml}$ and $\beta_{\rm sim}$ in Section 3 is limited to comparing the difference between mean values of $\beta_{\rm ml}$ and $\beta_{\rm sim}$ for each DCF equation, and to comparing the statistical and nonstatistical portions of each correction. These differences are discussed in Section 4.3.

The estimates of $\beta_{\rm ml}$ in Section 3 are derived from a normal distribution of angles as found by OSG01, and from assuming that the DCF numerator is either uniform, or nonuniform but uncorrelated with the angle dispersion in the denominator. The uncertainty in $\beta_{\rm ml}$ due to these assumptions has not been evaluated in detail. However, it may be estimated by testing the relation between mean column density $N$ and angle dispersion $\sigma_\theta$ for correlation, over a series of regions in a polarization map. Such tests can be applied to observations and to simulations. In Figure 5 of OSG01, the statistical distribution of $N$ becomes broader as the field strength becomes weaker and as the angle dispersion increases. Thus such simulations may have enough range in their variables to be tested for correlation between density and angle dispersion. One may expect the estimates of $\beta_{\rm ml}$ in Section 3 to be most accurate in map regions where the correlation between column density and angle dispersion is found to be weak or negligible.

The uncertainty in $\beta_{\rm ml}$ due to the assumption of negligible variation in the DCF equation numerator $\rho^{1/2}\sigma_v$ may also be tested using a suite of MHD simulations designed to sample increasing levels of turbulence, from nearly uniform clouds with small nonthermal pressure fluctuations and angle dispersions, to clouds with transonic turbulence, to clouds with supersonic turbulence. It may be possible to identify the level of fluctuations in $\rho^{1/2}\sigma_v$ where $B_{\rm ml}$ departs significantly from its value for uniform $\rho^{1/2}\sigma_v$.

The DCF equation numerator $(4\pi\rho)^{1/2}\sigma_{\mathrm{v}}$ is essentially the square root of the kinetic pressure. One may expect its fluctuations to be relatively small in cloud regions where the local pressure is in approximate balance with gravitational and magnetic forces. This requirement may be met for clouds which are close to their critical virial mass including external pressure and magnetic fields (McKee 1999, Li et al. 2022). An additional requirement is that local pressure fluctuations are not significantly amplified by local stellar feedback in the form of outflows, radiative heating, and ionizing radiation.

**4.3. Implications**

In Section 3, a statistical interpretation indicates that the most likely value of the local DCF field strength is always less than the global value expressed in the DCF equation. This result $\beta_{\mathrm{ml}} < 1$ applies to all versions of the DCF equation, provided their numerator values $\rho^{1/2}\sigma_{\mathrm{v}}$ are sufficiently uniform, and that their POS polarization angles have a normal distribution. This statistical bias arises because reciprocal normal distributions and related distributions have modal values less than unity (Johnson et al. 1994) as discussed in Section 3.2.

Table 2 shows that the typical statistical correction factor $\beta_{\mathrm{ml}}$ from $B_{\mathrm{DCF}}$ to $B_{\mathrm{ml}}$ is similar to the typical simulation correction factor $\beta_{\mathrm{sim}}$ from $B_{\mathrm{DCF}}$ to $B_{\mathrm{sim}}$, since the mean value of $\Delta\beta$ over all five DCF equations is close to zero. However this similarity in mean values is due in part to DCF equations where $\beta_{\mathrm{ml}}$ lies significantly below or above $\beta_{\mathrm{sim}}$ (e.g. $\Delta\beta = -0.3$ for the parallel $\delta B$ equation and $\Delta\beta = 0.4$ for the large-angle equation).

It is more realistic to interpret $\beta_{\mathrm{sim}}$ as arising from a combination of statistical and physical causes. Then values of the relative statistical correction $\gamma_{\mathrm{ml}}$ and the difference of mean correction factors $\Delta\beta$ in Table 2 can be used to rank the DCF equations according to the relative importance of their statistical and physical corrections. The equations are ranked by assuming that equations requiring the least physical correction are the most accurate estimators of field strength. For this purpose it is simplest to compare values of the relative physical correction $\gamma_{\mathrm{phys}} = 1 - \gamma_{\mathrm{ml}}$.

Three DCF equations in Table 2 have small relative physical corrections $\gamma_{\mathrm{phys}} = 0.1, 0.3,$ and $0.4$. They are the standard DCF equation with $\gamma_{\mathrm{phys}} = 0.1$ and $\Delta\beta = 0$, the structure function equation with $\gamma_{\mathrm{phys}} = 0.3$ and $\Delta\beta = 0.1$, and the original DCF equation with $\gamma_{\mathrm{phys}} =$ 0.4 and $\Delta\beta = 0.2$. They require relatively little physical correction, likely due to a combination of the causes described in Section 1.

In contrast two DCF equations have larger relative physical corrections and larger correction factor differences. The large-angle DCF equation has $\gamma_{\rm phys} = 0.6$ and $\Delta\beta = 0.4$. This correction is a reduction in field strength from the statistically corrected value of $B_{\rm ml}$. Such a reduction is most often attributed to Alfvén waves which have less energy than needed for equipartition with their turbulent driving (Falceta-Gonçalves et al. 2008, Liu et al. 2021). Other physical explanations are also described in Section 1.

The parallel-$\delta B$ equation has $\gamma_{\rm phys} = 0.7$ and $\Delta\beta = -0.3$. The required physical correction is an increase from the statistically corrected value of $B_{\rm ml}$, in contrast to the decrease from $B_{\rm ml}$ for the Alfvénic equations discussed above. This increase has been attributed to integration of 3D field vectors along the line of sight, which increases the effective angle dispersion in the plane of the sky (Skalidis & Tassis 2021; however, see C22 Section 5).

Following the above comparisons, the DCF equation which requires the least physical correction is summarized here. The standard DCF equation with $\gamma_{\rm phys} = 0.1$ and $\Delta\beta = 0$ is

$$B_{\rm ml,s} = \overline{\beta_{\rm ml,s}} B_{\rm DCF,s} = \overline{\beta_{\rm ml,s}} (4\pi\rho)^{1/2}\, \sigma_v / \tan\sigma_\theta \tag{14}$$

where

$$\overline{\beta_{\rm ml,s}} \approx \overline{\beta_{\rm sim,s}} \approx 0.7 \tag{15}$$

and where $\overline{\beta_{\rm ml,s}}$ and $\overline{\beta_{\rm sim,s}}$ are obtained from Table 2. The uncertainty in $\beta_{\rm ml,s}$ due to measurement uncertainty $\delta\theta$ is obtained from equation (12) as

$$\delta\beta_{\rm ml,s} \approx (1/2)(\delta\theta/\sigma_\theta). \tag{16}$$

It has long been unclear why DCF correction factors derived from simulations differ from one simulation to the next and from one DCF equation to the next. This variation has been discussed in terms of several proposed physical and geometrical explanations as noted in Section 1. This paper shows that for three DCF equations, a statistical correction can account for a significant portion of the typical correction needed to match simulation results, so that a relatively small physical correction is required. However, a significant physical correction is still required for two DCF equations, even after accounting for the most-likely statistical correction. This remains to be addressed in future simulation studies.

## 5. Summary and Conclusions

This section summarizes the main points of this paper and presents its main conclusions.

1. Five DCF equations which estimate the plane-of-the-sky field strength $B_{\mathrm{DCF}}$ are analyzed to obtain their corresponding "most-likely" field strength as proposed by C22. The probability distribution of the ratio $\beta$ of local to global field strengths is derived for each equation.

2. The most likely value of each distribution $p(\beta)$ is the modal value $\beta_{\mathrm{ml}}$ obtained by differentiation of $p(\beta)$. The most-likely plane-of-the-sky field strength is then $B_{\mathrm{ml}} = \beta_{\mathrm{ml}} B_{\mathrm{DCF}}$ where $\beta_{\mathrm{ml}}$ is the "most-likely correction factor" from $B_{\mathrm{DCF}}$ to $B_{\mathrm{ml}}$.

3. The values of $\beta_{\mathrm{ml}}$ are set primarily by the dependence of the DCF equation on angle dispersion $\sigma_\theta$. When $B_{\mathrm{DCF}} \propto \sigma_\theta^{-q}$, the mode of the probability distribution $p(\beta)$ occurs at $\beta_{\mathrm{ml}} = (1+q)^{-q/2}$. Four DCF equations based on excitation of Alfvénic MHD modes perpendicular to the mean field direction have $\beta_{\mathrm{ml}}$ close to its value for $q = 1$, for small angles. For these equations $\beta_{\mathrm{ml}} \approx 2^{-1/2} = 0.71$, the modal value of the reciprocal normal distribution (Johnson et al. 1994). For the parallel-$\delta B$ equation based on non-Alfvénic modes, $q = 1/2$ and $\beta_{\mathrm{ml}} = (3/2)^{-1/4} = 0.90$.

4. The uncertainty in $\beta_{\mathrm{ml}}$ due to angle measurement uncertainty $\delta\theta$ is estimated by propagation of errors as $\delta\beta_{\mathrm{ml}} = g(q)\frac{\delta\theta}{\sigma_\theta}$ where $g(q) \equiv q(1+q)^{-(1+q)/2}$, for $q = 1$ or $q = 1/2$ as in #3 above.

5. The values of $\beta_{\mathrm{ml}}$ in Table 2 vary by a factor of ~1.3 from 0.7 to 0.9. These values are set by the properties of the DCF equations as described in Section 3.2. In contrast the mean simulation correction factors vary by a factor of ~4, from ~0.3 to ~1.2. The relatively small range of $\beta_{\mathrm{ml}}$ values is consistent with the suggestion that the dispersion of polarization angles is not the main source of uncertainty in the DCF method (C22).

6. The most-likely correction factors and simulation correction factors differ by $\Delta\beta \leq 0.2$ for each of the original, standard and structure function equations. However they differ by 0.4 for the large-angle equation, and by -0.3 for the parallel-$\delta B$ equation. Thus values of $\beta_{\rm ml}$ alone cannot account for the values of $\beta_{\rm sim}$.

7. It is more useful to assume that each "correction" in field strength from a DCF estimate to a simulation value is due to a combination of statistical and physical corrections. For each DCF equation the relative importance of the physical correction is evaluated by the "relative physical correction" $\gamma_{\rm phys}$, with possible values from 0 to 1. The standard, structure function, and original DCF equations are ranked as the most accurate estimators of field strength, with $\gamma_{\rm phys} = 0.1$, 0.3, and 0.4. On the other hand the parallel-$\delta B$ and large-angle equations appear less accurate, with $\gamma_{\rm phys} = 0.6$ and 0.7.

8. These conclusions are limited by simplifying assumptions of the cloud map model to low inclination of the mean field direction $\gamma \lesssim 30°$, low angle dispersion in the plane of the sky $\sigma_\theta \lesssim 30°$, and by relatively small fluctuations in turbulent cloud pressure across the map. They may also be limited by the diversity of published simulation correction factors used in the analysis.

9. These results may help to select the most useful DCF equations for analysis of polarization observations. In future studies, they may be useful to clarify the leading physical mechanisms of DCF equation correction.

**Acknowledgements**

We thank Che-Yu Chen and Zhi-Yun Li for their paper C22, which inspired us to undertake this project, and for their helpful comments and suggestions. We thank the reviewer for comments which helped to clarify and improve the paper. This work was developed as part of the NASA/DLR Stratospheric Observatory for Infrared Astronomy (SOFIA) FIELDMAPS Legacy program. SOFIA is jointly operated by the Universities Space Research Association, Inc. (USRA), under NASA contract NNA17B53C, and the Deutsches SOFIA Institut (DSI) under DLR contract 50 OK 0901 to the University of Stuttgart. IWS and SC acknowledge financial support provided by NASA through award #08_0186 issued by USRA.